\documentclass[aip,graphicx]{revtex4-1}
\usepackage{bm}
\usepackage{amsmath}
\usepackage{amssymb}
\usepackage{graphicx}
\usepackage{color}

\draft 

\begin{document}

\title{Response to ``Comment on `Faraday waves in a Hele-Shaw cell' [Phys Fluids 30, 042106 (2018)]"} 

\author{Jing Li}
\email[]{lijing\_@sjtu.edu.cn}
\affiliation{Marine Numerical Experiment Center, State Key Laboratory of Ocean Engineering, Shanghai 200240, China}
\affiliation{School of Naval Architecture, Ocean and Civil Engineering, Shanghai Jiao Tong University, Shanghai 200240, China}
\author{Xiaochen Li}
\affiliation{School of Civil Engineering and Transportation, South China University of Technology, Guangzhou 510641, China}
\author{Shijun Liao}
\affiliation{Marine Numerical Experiment Center, State Key Laboratory of Ocean Engineering, Shanghai 200240, China}
\affiliation{School of Naval Architecture, Ocean and Civil Engineering, Shanghai Jiao Tong University, Shanghai 200240, China}

\date{\today}


\maketitle 

In their comment\cite{boschan2023}, Boschan \emph{et al.} checked the scaling law proposed in Ref.\cite{li2018}.
Both their experiments and ours yield a similar result, which again validates the dimensionless relation obtained from Buckingham's method in Ref.\cite{li2018}.
However, they proposed a new scaling law in that comment based on two arguments: First, the wave height $H$ is independent of the liquid depth $d$ when $d \ge 10$mm which is the parameter range in Ref.\cite{li2018}; Second, Faraday waves are triggered only if the acceleration amplitude $F$ exceeds a critical value $F_c$.

\begin{figure}
\centering
\includegraphics[width=0.6\textwidth]{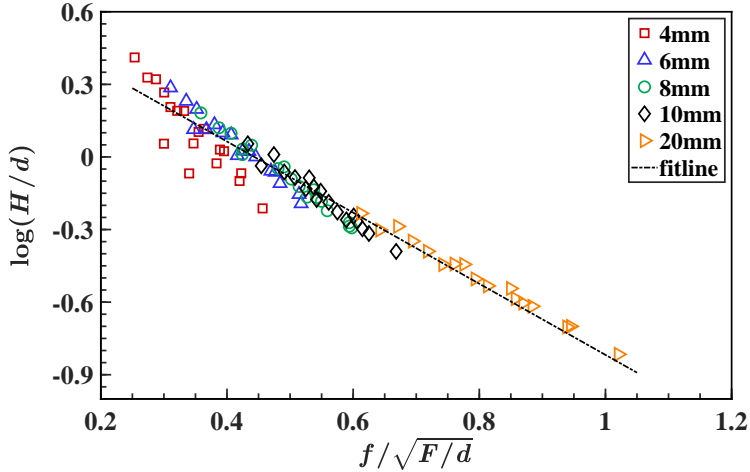}
\caption{\label{fig:scale}Diagram of experimental data averaged from two parallel experiments with the same oscillating parameters. Different symbols indicate data at corresponding liquid depths and the line is fitted from them. The Hele-Shaw cell size is 300mm$\times$1.5mm$\times$50mm. The frequency $f$ is from 15 to 25Hz, and the acceleration amplitude $F$ is from 10 to 18m/s$^2$.}
\end{figure}

\begin{figure}
\centering
\includegraphics[width=0.6\textwidth]{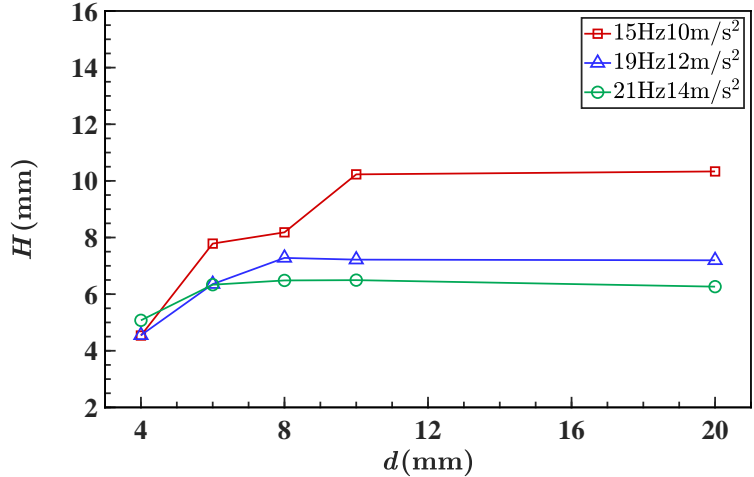}
\caption{\label{fig:depth}Data of $H$ against $d$ on three oscillating conditions of the present experiments, averaged from two parallel experiments with the same oscillating parameters.}
\end{figure}

The dimensionless variable $f/\sqrt{F/d}$, a variant of the Strouhal number, obtained from Buckingham's method rather than from experiments not only applies to cases in Ref.\cite{li2018} but also to more general problems at different liquid depths, though limited.
Hence, the logarithmic relation should be valid over a wide parameter regime where $H$ is dependent on $d$. 
To demonstrate this, we add a series of experiments, and the data is shown in Fig.\ref{fig:scale} where $d$ encompasses the range of 4 to 20mm. 
As the comment mentioned, in our previous work (Fig.5 in Ref.\cite{li2015}) $H$ is independent of $d$ when $d \ge 10$mm and this number can be as low as 6mm in other studies (cited in their comment). 
The dependence of $H$ on $d$ is hence shown in Fig.\ref{fig:depth} with respect to three combinations of frequency and acceleration from the present experiments.
So far, it is clear that $d$ is a key variable for this problem as can be seen from Fig.\ref{fig:depth}, and the good match of our results in Fig.\ref{fig:scale} shows the correctness of the scaling law 
\begin{equation}
    \log(\frac{H}{d})\approx -\alpha\frac{f}{\sqrt{F/d}}+\beta,
\end{equation}
where coefficients $\alpha$ and $\beta$ vary for differing circumstances according to liquid property, container material, gap size and etc.

As for the second argument, finding a simple and feasible model to predict wave height is a significant concern.
Although the more factors taken into consideration, the closer the model will be, if the critical value $F_c$ is brought in, presumably one cannot obtain a priori information on the wave height, since $F_c$ is hard to know without laboratory measurements for Faraday instability in Hele-Shaw cells.
Actually, how to access $F_c$ theoretically is a cutting-edge problem for this type of container with a porous nature.
Besides, we have stressed the capillary effect in Ref.\cite{li2019stability}.
Given a similar weight in the dispersion relation, gravity and surface tension should be taken into account at the same time, but only gravity was involved in the comment.
The scaling law proposed in Boschan \emph{et al.}'s comment should thus be further improved by paying close attention to both these two factors. 

Summarizing, we stress the correctness of including liquid depth $d$ as a key parameter in dimensional analysis for this problem and the practical feasibility of the simple form of the scaling law.
Besides, $f/\sqrt{F/d}$ is a variant of the Strouhal number which is a widely known and physically correct dimensionless variable for assessing oscillating flow.

\section*{Declaration of Interests}
The authors report no conflict of interest.

\section*{Data Availability}
The data that supports the findings of this study are available within the article and from the corresponding author upon reasonable request.

\end{document}